\documentclass[sigconf]{acmart}
\usepackage{import}
\usepackage[inline,shortlabels]{enumitem}
\usepackage[T1]{fontenc}

\AtBeginDocument{%
  \providecommand\BibTeX{{%
    \normalfont B\kern-0.5em{\scshape i\kern-0.25em b}\kern-0.8em\TeX}}}

\newcommand{\proj}{F-FADE\xspace}

\newcommand{\pan}[1]{\textcolor{blue}{(Pan: #1)}}
\newcommand{\panred}[1]{\textcolor{blue}{(Pan's comment: #1)}}

\newcommand{\todo}[1]{\textcolor{red}{(TODO: #1)}}
\newcommand{\hide}[1]{}

\theoremstyle{definition}
\newtheorem{theorem}{Theorem}[section]
\newtheorem{proposition}[theorem]{Proposition}
\newtheorem{definition}[theorem]{Definition}

\newtheorem{lemma}[theorem]{Lemma}

\theoremstyle{remark}

\usepackage{multirow}
\usepackage[linesnumbered,ruled,lined,commentsnumbered]{algorithm2e}
\newcommand{\xhdr}[1]{\vspace{2mm}{\noindent\bfseries #1}.}

\usepackage[absolute,overlay]{textpos}
\usepackage{subfig}

\setcopyright{acmcopyright}
\copyrightyear{2021}
\acmYear{2021}
\acmDOI{10.1145/3437963.3441806}

\acmConference[WSDM '21] {The 14th ACM International Conference on Web Search and Data Mining}{March 8--12, 2021}{Virtual Event, Israel}
\acmBooktitle{The 14th ACM International Conference on Web Search and Data Mining (WSDM '21), March 8--12, 2021, Virtual Event, Israel}
\acmPrice{15.00}
\acmISBN{978-1-4503-8297-7/21/03}

\begin{document}
\fancyhead{}
\title{\proj: Frequency~Factorization~for~Anomaly~Detection in~Edge~Streams}

\author{Yen-Yu Chang}
\authornote{Yen-Yu Chang and Pan Li contributed equally to this research.}

\affiliation{%
  \institution{Stanford University}
}
\email{yenyu@cs.stanford.edu}

\author{Pan Li}
   \authornotemark[1]
\affiliation{%
  \institution{Purdue University}
}
\email{panli@purdue.edu}

\author{Rok Sosic}
\affiliation{%
  \institution{Stanford University}
}
\email{rok@cs.stanford.edu}

\author{M. H. Afifi}
\affiliation{%
 \institution{Barracuda Networks}
}
\email{mibrahim@barracuda.com}

\author{Marco Schweighauser}
\affiliation{%
  \institution{Barracuda Networks}
}
\email{mschweighauser@barracuda.com}

\author{Jure Leskovec}
\affiliation{%
  \institution{Stanford University}
}
\email{jure@cs.stanford.edu}

\begin{abstract}
Edge streams are commonly used to capture interactions in dynamic networks, such as email, social, or computer networks.
The problem of detecting anomalies or rare events in edge streams has a wide range of applications.
However, it presents many challenges due to lack of labels, a highly dynamic nature of interactions, and the entanglement of temporal and structural changes in the network. Current methods are limited in their ability to address the above challenges and to efficiently process a large number of interactions.
Here, we propose \proj, a new approach for detection of anomalies in edge streams, which uses a novel frequency-factorization technique to efficiently model the time-evolving distributions of frequencies of interactions between node-pairs. The anomalies are then determined based on the likelihood of the observed frequency of each incoming interaction. \proj is able to handle in an online streaming setting a broad variety of anomalies with temporal and structural changes, while requiring only constant memory.
Our experiments on one synthetic and six real-world dynamic networks show that \proj achieves state of the art performance and may detect anomalies that previous methods are unable to find.
\end{abstract}

\vspace{-2mm}
\vspace{-2mm}

\maketitle

\vspace{-2mm}
\section{Introduction}
\vspace{-0.2mm}
\begin{figure*}
\includegraphics[trim=0.2cm 13.1cm 0.3cm 4.2cm, clip, width=0.9\textwidth]{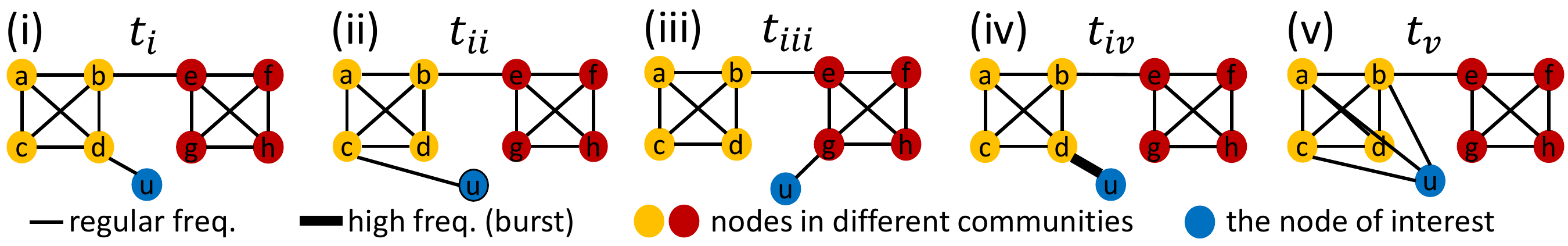}
\begin{tabular}{|c|c|c|c|}
\hline 
Compare patterns & Condition &  Behavior type (anomaly indication) & More indicative type of information  \\
\hline
(ii) vs. (i) & $t_{ii} > t_{i}$ & Interactions within the group (no anomaly) &  Structural info. \\
\hline 
(iii) vs. (ii) & $t_{iii} = t_{ii}$ & Prompt group change (anomaly) & Structural + Temporal info.  \\
\hline
(iii) vs. (ii) & $t_{iii} \gg t_{ii}$ & Long-term group change (unknown) & Structural + Temporal info. \\
\hline
(iv) vs. (ii) & $t_{iv} = t_{ii}$ & Burst of interactions (anomaly) &  Temporal info. \\
\hline
(v) vs. (ii) & $t_{v} = t_{ii}$ & Burst of interactions with a group (anomaly) & Structural + Temporal info. \\
\hline
\end{tabular}
\vspace{-3mm}
\caption{Patterns of interactions in a dynamic network. (i) is an initial interaction, known to be regular, of an external node $u$ with a group member $d$. Later interactions of $u$ can be: (ii) with a different node from the same group, (iii) with a node from a different group, (iv) with the same node, but at a much higher frequency, (v) with nodes from the same group, where all pairwise interactions are regular, but group level interactions are at an increased frequency. The table shows for each pattern which ones are most likely to be anomalous and what types of information are needed to identify the anomalies.}\vspace{-1mm}
\label{fig:pattern}
\end{figure*}

An edge stream refers to the time-ordered sequence of edges in a dynamic network, a commonly used representation of complex systems~\cite{carley2003dynamic}. Edges typically correspond to dyadic interactions in those systems. For example, in an email network, the edge stream consists of time-ordered emails that record interactions from senders to recipients, representing the dynamic communication network between users~\cite{shetty2004enron}. Thus, edges and interactions will be used interchangeably later on. 
The goal of anomaly detection in edge streams is to find unusual edges. These can identify important undesirable activity in the system.
In the case of email networks, regular users can be harmed by malicious messages, such as phishing, or compromised accounts~\cite{jagatic2007social,hu2016detecting}. In transaction networks, anomalous transactions can indicate financial fraud or money laundering~\cite{penzias1994fraud}. 

As anomalies are caused by rare events by definition, it might be impossible to find sufficient number of training labels for supervised methods. We thus focus here on unsupervised anomaly detection approaches. Most current approaches are snapshot-based where edge streams are aggregated into network snapshots over time~\cite{ranshous2015anomaly}. These approaches can detect anomalies only after an entire snapshot has been collected, which can introduce a significant and often prohibitive time lag. Instead, we want anomalies in edge streams to be reported in an online, streaming fashion as soon as anomalous interactions happen. However, a streaming approach to anomaly detection in edge streams introduces two major challenges. 

First, the approach must be able to handle 
temporal and structural changes simultaneously. 
We illustrate this point by showing different patterns of anomalies in dynamic networks in Fig.~\ref{fig:pattern}. In case of pattern (iii), where an external node $u$ starts interacting with a node from a different group (from a yellow node $d$ to a red node $g$), we need to know that the nodes belong to different groups (a structural change) as well as the time between the interactions (a temporal change). The interaction is more likely anomalous, if the time is short due to the rapid switch to another group~\cite{aggarwal2012event,hu2016embedding}. Similarly, we need both temporal and structural changes in case of pattern (v), where $u$ interacts with many nodes from the same group. All pair-wise interactions between $u$ and other individual nodes can be regular, yet when viewed together, these interactions can show an anomalous increase in the group interaction frequency. 

Second, the approach should be time and memory efficient as a large number of interactions can take place in a short time, which significantly constrains the available time for analysis of each edge. This constraint is especially limiting in the case of streaming approaches that aim to digest both temporal and structural changes simultaneously and require significant time and space resources to compute. For example, an aggregation of timestamps for estimating time-evolving distributions of interactions must keep track of the network structural information, but a straightforward approach leads to an increasingly large memory cost as interactions between new pairs of nodes are encountered. Although snapshot-based approaches, by utilizing significant time and space resources, are able to analyze the entanglement of structural and temporal information, it remains an open problem of how to do that in a stream-based fashion.
In summary, we want an efficient, unsupervised method for detecting anomalies in edge streams, where the method works in a streaming manner and is able to take advantage of both temporal and structural information in order to detect a wide range of anomalous interaction patterns.

\xhdr{Present Work}\footnote{The code and supplements are available at \url{http://snap.stanford.edu/f-fade/}.}
Here, we propose \textbf{F}requency-\textbf{F}actorization for \textbf{A}nomaly \textbf{DE}tection (\proj), a new approach for detection of anomalies in edge streams. A key innovation of \proj is a novel frequency-factorization technique that is able to handle a broad variety of anomalies, such as those illustrated in Fig.~\ref{fig:pattern}, while requiring only constant memory.
Specifically, \proj models the time-evolving distributions of frequencies of interactions between nodes of a dynamic network and determines the anomalies based on the likelihood of the observed frequency of an incoming interaction.
Using an online factorization approach, \proj efficiently handles the structural information in order to estimate the latent parameters of the distributions thus reflecting the intensity of frequencies in a maximum likelihood rule.
Furthermore, \proj keeps in memory only the most frequent interactions, which results in constant memory use. Overall, our contributions are as follows:
    
i. As opposed to the previous probabilistic methods, \proj uses an efficient online factorization to fully incorporate the network structural information, and therefore can detect anomalies that are caused by rapid changes in the network structure.

ii. As opposed to previous matrix factorization approaches, our novel approach operates in the space of the intensity of interaction frequencies, corresponding to a statistical model that properly controls false positive rates under a mild assumption. This statistical model of interaction frequencies further allows \proj to detect anomalies in a group of simultaneous interactions rather than being limited to interactions between two nodes.

iii. \proj is an online method and has constant memory cost even with incoming new nodes. 
We are not aware of any previous matrix factorization approaches that exhibit this property.

iv. We evaluate \proj on the edge streams of three public real dynamic networks and four email networks of real companies. Our method significantly outperforms all the baselines and may detect anomalies that previous methods are unable to find.

\section{Preliminaries and Related Work} \label{sec:related}
\vspace{-0.2mm}
As a preliminary, we classify interactions in dynamic networks into a number of basic patterns as illustrated in Fig.~\ref{fig:pattern}, where two groups, i.e. organizations, are represented by yellow and red nodes, and their regular interactions form the base of the network. We are interested in finding out if any of the interactions of node $u$, which does not belong to any of the groups, are anomalous or regular. Given a regular interaction between $u$ and node $d$ from one of the organizations (Fig.~\ref{fig:pattern} (i)), then the follow-up interactions of $u$ can be assigned to one of four other patterns Fig.~\ref{fig:pattern} (ii)-(v).


Looking only at temporal or only at structural changes might be sufficient to identify anomalies for some of the above patterns. For example, in case of (ii), since $u$ interacts with another node from the same group, the interaction is most likely regular, as interacting with nodes of the same group is a common behavior in many real networks. We thus need to rely on group membership, a structural information type, to identify the group to which a node belongs. Similarly, in case of (iv), we need the frequency of interactions between two nodes, which is a purely temporal information type.

However, to identify anomalies in more complex patterns, we need to take into account both temporal and structural changes. In case of (iii), if this interaction is close in time to the initial interaction, then it is more likely anomalous due to the prompt switch to another group~\cite{aggarwal2012event,hu2016embedding}, which is detectable by leveraging structural information. On the other hand, if the interaction occurs much later, then temporal aspects become more critical since group membership is less informative after a long time.

As demonstrated by case (v), anomaly detection in the case of sudden changes in the frequency of interactions must also take structural information into account. Node $u$ interacts with nodes from the same group as previously and all pair-wise interactions between $u$ and other individual nodes appear to be regular, yet when viewed together, these interactions can turn out to be anomalous.
 
\vspace{-0.5mm}
\xhdr{Related Work} 
Many publications concern anomaly detection for dynamic networks~\cite{akoglu2015graph,ranshous2015anomaly,cadena2018graph}.
We briefly review them to discuss their methodological foundations and related limitations in detecting some anomaly patterns from Fig.~\ref{fig:pattern}.

\emph{Probabilistic methods} rely on probabilistic models that characterize the regular communication patterns of the dynamic network and determine anomalies based on the pattern deviation from the models. Probabilistic methods by nature allow computation of $p$-values (or false positive rates equivalently) of their detection even for a group of interactions~\cite{heard2010bayesian,aggarwal2011outlier,bhatia2020midas}. However, they either require a complex optimization over the entire graph by recording all historical data~\cite{priebe2005scan,aggarwal2011outlier,wang2013locality,neil2013scan,chen2014non,peel2015detecting} or only capture limited structural information restricted in a local network region~\cite{heard2010bayesian,yu2013anomalous,bhatia2020midas}. Specifically, the most recent probabilistic method on anomaly detection in edge streams~\cite{bhatia2020midas} is unable to track community structures and thus fails to differentiate patterns Fig.~\ref{fig:pattern} (ii) and (iii).


\emph{Matrix factorization methods}~\cite{sun2007less,teng2017anomaly,yu2017temporally,sun2006beyond} leverage the ``low-rank'' property of real-world network structures~\cite{mardani2012dynamic} that is mostly represented as overlapping, non-overlapping, or hierarchical community structures~\cite{fortunato2010community}. Anomalies break the low-rank property and are thus detectable. Matrix factorization approaches globally capture the structural information but they can neither control $p$-values of the anomaly detection nor detect a group of simultaneous interactions with proper calibration, such as the pattern Fig.~\ref{fig:pattern} (v). Moreover, to the best of our knowledge, no previous matrix factorization method works on edge streams, thus these approaches cannot handle new arriving nodes or provide a timely detection response. 

\emph{Distance-based methods} propose certain time-evolving measures of dynamic network structures and use the change rates of those measures to detect anomalies. These measures include PageRank~\cite{eswaran2018sedanspot,yoon2019fast}, the embeddings of nodes~\cite{yu2018netwalk} or entire networks~\cite{eswaran2018spotlight}, and other handcrafted features~\cite{wang2015localizing,ranshous2016scalable,henderson2010metric}. However, these methods present additional limitations besides the loss of control in $p$-values of their detection. For example, SedanSpot~\cite{eswaran2018sedanspot} cannot detect the change from  the pattern (i) to patterns (iv) and (v) in Fig.~\ref{fig:pattern} because the personalized PageRank~\cite{jeh2003scaling} that SedanSpot tries to approximate is hardly identifiable by nature. AnomRank~\cite{yoon2019fast} introduces two specifically designed transportation vectors to address this issue, but needs to compute a global PageRank, which does not scale for edge stream processing. Node embeddings given by the auto-encoder~\cite{yu2018netwalk} may also have unidentifiable changes from the pattern (i) to patterns (iv) and (v) in Fig.~\ref{fig:pattern}. 

Some recent works may process attributed networks~\cite{ding2019deep,manzoor2016fast,li2017radar,shah2016edgecentric}, which is not our focus. However, it is interesting to study in the future whether our key technique, frequency factorization (introduced later), can be applied there. Other works on counting persistent patterns may be used for burst and periodic anomalies (iv)~\cite{belth2020mining}, while they cannot detect structural anomalies (iii) and (v).  

\vspace{-0.5mm}
\section{Problem Formulation and Notation} 
\vspace{-0.2mm}
Let $\mathcal{E} =\{e_1, e_2, e_3,...\}$ be a stream of interactions from a dynamic network. Each $e_i$ is a 4-tuple $(s_i, d_i, t_i, w_i)$, where 
$s_i$ and $d_i$ are the source and the destination nodes, respectively,
$t_i$ is the interaction time, and $w_i$ is the interaction count. We call the pair of the source and destination node $(s_i$, $d_i$) to be \emph{the interaction type}. 
Without loss of generality, we can assume that $t_i$ is represented as a positive integer whose unit reflects the systematic time granularity. 

\vspace{-0.7mm}
\xhdr{Problem Formulation} Our task is to detect anomalous interactions in $\mathcal{E}$ which could belong to any of the patterns shown in Fig.~\ref{fig:pattern}. Specifically, the method is expected to utilize temporal and structural information to detect interactions that impose either a prompt change of the network structure, belong to the burst of interactions with a single node or a group of nodes as shown in Fig.~\ref{fig:pattern} (iii)-(v).
Moreover, the method is expected be online and capable of processing large amounts of data with bounded memory. 

\vspace{-0.5mm}
\xhdr{Notation} We introduce \emph{interaction-temporal-frequency map} (ITFM) and \emph{interaction-type set} (ITS), two frequently used data structures.
\vspace{-1mm} 
\begin{definition}
The \emph{interaction-temporal-frequency map} (ITFM) $\langle(s,d),(t,f)\rangle$ maps each interaction type $(s,\,d)$ to $(t,\,f)$, where $t$ is a time stamp (a positive integer) and $f$ denotes frequency (a real value). The \emph{interaction-type set} (ITS) is a set of interaction types, i.e., the keys of an ITFM, denoted by  $\{(s,d)\}$. We define the operation \emph{ITS($\cdot$)} which transforms one ITFM into the corresponding ITS. \vspace{-1mm}
\end{definition}
For ITMF $F$, we use $F(s,d)$ to denote the mapping of the key $(s,d)$ to its corresponding $(t,\,f)$ in $F$.
An ITFM or an ITS can be viewed as directed graphs between the nodes, with ITMFs having additional edge attributes. We define the node set based on $F$ as $V(F)=\cup_{(s,d)\in F}\{s,d\}$. We also define the in and out neighbors of a node $v\in V(F)$ as $\mathbf{N}_{\text{in}}(v, F) = \cup_{s:(s,v)\in F}\{s\}$ and $\mathbf{N}_{\text{out}}(v, F) = \cup_{d:(v,d)\in F}\{d\}$ respectively. Finally, let $\mathcal{N}(0,1)$ denote the standard normal distribution and let $\mathbb{P}(\cdot)$ denote a probability distribution. We may maintain \emph{node embeddings} $H$ that associates each node $v\in V(F)$  with an $m$-dimensional vector $h_v\in\mathbb{R}^{m}$.
\vspace{-0.8mm}
\section{Method}
\vspace{-0.2mm}
\begin{algorithm}[t]
\SetKwInOut{Param}{Param}\SetKwInOut{Input}{Input}\SetKwInOut{Output}{Output}\SetKwInOut{Setup}{Setup}
\SetKwFunction{InteractionAgg}{AGG} 
\SetKwFunction{Eval}{DETECT}
\SetKwFunction{Union}{UNION}
\SetKwFunction{Update}{F-FAC} 
\Input{Edge stream $\mathcal{E}$; Param.: $t_{\text{setup}}$, $W_{\text{upd}}$, $\alpha$, $M$, $m$, $f_{\text{th}}$}
\Output{An anomaly score stream $\text{Sc}^{(t)}$, $t=t_{\text{setup}}+1, ...$}
$k\leftarrow 0$, $\text{Act-S},F,H\leftarrow \emptyset$, $Q\in\mathbb{R}^{m\times m}$ where $Q_{ij}\stackrel{\text{iid}}{\sim}\mathcal{N}(0,1)$\;
\For{$e\leftarrow (s,d,t,w)$ in $\mathcal{E}$,}{
\lIf{$t>t_{\text{setup}}$}{$\text{Sc}^{(t)} \leftarrow$ \Eval{$F$, $e$, $H$, $Q$, $f_{\text{th}}$}}
$F$, $\text{Act-S}$, $f_{\text{th}} \leftarrow$ \Union{$F$, $\text{Act-S}$, $e$, $\alpha$, $M$, $t$}\;
\If{$t \geq t_{\text{setup}} + kW_{\text{upd}}$}{
$H \leftarrow$ \Update{$F$, $\text{Act-S}$, $H$, $Q$, $f_{\text{th}}$}\;
$k\leftarrow k+1$, $\text{Act-S}\leftarrow \emptyset$\;
}
}
\caption{\proj ($\mathcal{E}$, $t_{\text{setup}}$, $W_{\text{upd}}$, $\alpha$, $M$, $m$, $f_{\text{th}}$)}
\label{alg:F-FADE}
\end{algorithm}
\begin{table}[t]
        \vspace{-8mm}
\end{table}

\begin{table}[t]
    \centering
    \vspace{-3mm}
    \begin{tabular}{|c|l|}
    \hline
    $t_{\text{setup}}$ & The time to set up the model  \\
    \hline
    $W_{\text{upd}}$ & The time interval for model update, integers \\
        \hline
    $\alpha$ & The decay rate when updating frequency, range $[0,1)$  \\
    \hline
    $M$ & The upper limit of memory size \\
    \hline
    $m$ & The dimension of node embeddings \\
         \hline
    $F$ & An ITFM to record interactions with their frequencies\\
    \hline
    $H$ & The embeddings of active nodes \\
        \hline
    $f_{\text{th}}$ & The cut-off threshold of the frequency to record \\
        \hline
    $Q$ & A random full rank matrix used in our model (Eq.~\eqref{eq:ffobj})\\
             \hline
     $\text{ACT-S}$ & An ITS to record active interaction-types  \\
         \hline
    \end{tabular}
    \caption{Variables in \proj}
    \label{tab:variables}
        \vspace{-7.5mm}
\end{table}

In this section, we introduce our proposed approach \proj, shown in Alg.~\ref{alg:F-FADE} (Table~\ref{tab:variables} provides a description of variables). \proj includes three key components. First, \proj maintains an ITFM $F$ that consists of a bounded number of node-pairs with temporarily high-frequent interactions between them.
$F$ essentially records the network skeleton and keeps updated when new interactions come in (the subroutine UNION). Second, after an initial short setup period from 0 to $t_{\text{setup}}$, which is needed to establish $F$, 
node embeddings $H$ are learnt for every time window $W_{\text{upd}}$ via the frequency-factorization approach.
Note that we introduce an ITS ($\text{Act-S}$) to record the temporarily active types of interactions which allows for efficiently local update of node embeddings. 
These node embeddings parameterize the time-evolving distributions of interaction frequencies (the subroutine F-FAC). Third, for each new arriving interaction, an anomaly score will be assigned based on the likelihood of its observed frequency with respect to the distribution parameterized by node embeddings (the subroutine DETECT). 

Here, we first describe an online approach to efficiently maintain ITFM $F$, the network structure. In the next two subsections, we focus on the other two subroutines F-FAC and DETECT, respectively.  

At a certain time $t_i$, an element $\langle(s,d), (t,f)\rangle$ in $F$ indicates that the $(s,d)$-type interaction appeared before $t_i$ most recently at time $t$ and that the aggregated frequency of this interaction type is $f$. In general, the time-evolving aggregated frequency $f$ for $(s,d)$-type interactions at $t$ is computed as: \vspace{-1mm}
\begin{align}\label{eq:agg-freq}
\text{Agg-freq:}\quad\quad f\;\triangleq \sum_{(s,d,t',w)\in \mathcal{E}:t'<t} w*\text{ker}(t-t'),
\end{align}
where $\text{ker}(\cdot)$ is a kernel function for interaction aggregation. $\text{ker}(\cdot)$ is defined over $\mathbb{Z}_{\geq 0}$ and satisfies $\sum_{i=0}^\infty \text{ker}(i) = 1$. In this work, we set $\text{ker}(i) = \alpha^{i}(1-\alpha)$ for some $\alpha\in(0,1)$ and thus smaller $\alpha$ emphasizes more recent observed interactions. \proj maintains $F$ (line 4) to merge $e$ into $F$ (lines 2-3 of UNION). The parameter $M$ controls the size of $F$ which further controls the memory cost of \proj. Infrequent interactions will be removed and the corresponding cut-off frequency is recorded by $f_{\text{th}}$ (lines 4-7 of UNION).

\begin{algorithm}[t]
\SetKwInOut{Input}{Input}\SetKwInOut{Output}{Output}
\Input{An ITFM $F$, an ITS $\text{Act-S}$, $e$; Param.: $\alpha$, $M$, $t_0$}
\Output{The updated $F$, $\text{Act-S}$, the cut-off frequency $f_{\text{th}}$} 
Insert $(s(e), d(e))$ into $\text{Act-S}$\;
\lIf{$(s(e), d(e))$ is in $F$}{
$(t',f')\,\leftarrow \, F(s(e), d(e))$,
$F(s(e), d(e)) \,\leftarrow\,(t(e),\alpha^{(t(e)-t')}f' + (1-\alpha) w(e))$
}\lElse{ Insert $((s(e),d(e)),(t(e), (1-\alpha) w(e)))$ into $F$}
$f_{\text{th}}\leftarrow \min f' \,\, \text{s.t.}\,\, |\{\langle(s,d),(t,f)\rangle\in F|\alpha^{t_0-t}f\geq f'\}|\leq M$ \;
\For{$e=\langle(s,d),(t,f)\rangle\in F$, s.t. $\alpha^{t_0-t}f<f_{\text{th}}$}{Remove $e$ from $F$; Remove $(s,d)$ from \text{Act-S}\;}
\caption{UNION($F$, $\text{Act-S}$, $e$, $\alpha$, $M$, $t_0$)}\label{alg:union}
\vspace{0mm}
\end{algorithm}
\begin{table}[t]
        \vspace{-9mm}
\end{table}

\vspace{-0.8mm}
\subsection{Frequency Factorization}\label{sec:FF}
\vspace{-0.2mm}

\begin{algorithm}[t]
\SetKwInOut{Input}{Input}\SetKwInOut{Output}{Output}
\Input{An ITFM $F$, an ITS \text{Act-S}, node embeddings $H$; Param.: $Q$, $f_{\text{th}}$}
\Output{Updated node embeddings $H$} 
\lFor{$h_v\in H$, $v\notin V(F)$}{Remove $h_v$ from $H$}
\lFor{$v\in V(F)$, $h_v\notin H$}{Randomly initialized $h_v\in\mathbb{R}^{m}$}
\For{$\text{gradient ascent steps}=1,2,...$}{
\lIf{global optimization (at $t_{\text{setup}}$)}{Sample a mini-batch $\Omega\subseteq V(\text{Act-S})\times V(\text{Act-S})$}
\If{local optimization  (at $t_{\text{setup}} + kW_{\text{upd}}$, $k\geq 1$)}{
Sample a mini-batch $\Omega_p \subseteq \text{Act-S}$\;
Sample a mini-batch $V' \subseteq V(F)/V(\text{Act-S})$\;
$\Omega \leftarrow (V(\Omega_p)\cup V')\times (V(\Omega_p)\cup V') $
}

Do one-step gradient ascent over $\{h_v| v\in V(\text{Act-S})\}$ to increase 
 $\sum_{(s,d)\in \Omega} \log \mathbb{P}(f;\lambda_{sd})$, where $f=f_{sd}$ if $\langle (s,d),(t,f_{sd})\rangle \in F$ for some $t$ or $f=f_{\text{th}}$ otherwise\;
}
\caption{F-FAC($F$, $\text{Act-S}$, $H$, $Q$, $f_{\text{th}}$)}\label{alg:ffac}
\end{algorithm}
\begin{table}
\vspace{-10.5mm}
\end{table} 

Our approach is to maintain the time-evolving distributions of frequencies of interactions under regular node behavior. Then we observe the frequency of incoming interactions and use the likelihood-based model to determine whether they are anomalies or not. However, this approach presents a major challenge. In real networks, interactions between pairs of nodes are typically sparse. Even worse, the bounded memory cost allows us to track only the skeleton network structure consisting of highly frequent interactions. Therefore, if we determine the distribution of the interaction frequency between two nodes by only tracking their historical interaction frequency, the model will not be able to make good estimates when only a few or even no historical interactions are present. 

Our solution is to utilize network structures to address this limitation.
In general, a real network typically holds certain low-rank properties, which indicate that latent features of nodes may be extracted by factorizing the low-rank approximation of the adjacency matrix that represents the network skeleton. These properties have been widely used in many network-related applications, such as community detection~\cite{fortunato2010community,yang2013overlapping}, link prediction~\cite{dunlavy2011temporal}, recommendation system design~\cite{paterek2007improving} and also anomaly detection reviewed in Sec.~\ref{sec:related}. However, in contrast to previous factorization-based approaches, our approach is based on the max-likelihood rule to estimate the latent intensity parameters of the interaction frequencies.



Specifically, consider a probabilistic distribution of frequency $f$ with a single positive parameter $\lambda$, denoted by $\mathbb{P}(f;\lambda)$. Suppose the expectation $\mathbb{E}[f]$ monotonically increases with respect to $\lambda$ and thus $\lambda$ reflects the intensity of $f$.  One general class of such distributions is Gamma distribution: $\mathbb{P}(f;\lambda) = \frac{1}{\lambda^\theta \Gamma(\theta)}f^{\theta-1}\exp(-\frac{f}{\lambda})$,
for any $\theta>0$. In this work, we choose $\theta=1$, which corresponds to the exponential distribution and which performs well as we show later.

\xhdr{Frequency Model} Recall that we collect and summarize the aggregated frequencies of different interaction-types into the ITFM $F = \{\langle (s,d),(t,f)\rangle\}$. We associate each node in $V(F)$ with an embedding vector $h_v\in\mathbb{R}^{m}$ that changes over time. Then, our frequency model assumes that the frequency of interactions between two nodes, say $s$ and $d$, denoted by $f_{sd}$, follows the distribution: \vspace{-0.5mm}
\begin{align}\label{eq:f-model}
\quad  f_{sd} \sim \mathbb{P}(f;\lambda_{sd})= \exp(-f/\lambda_{sd})/\lambda_{sd}
\end{align}
where $\lambda_{sd} = \exp(h_s^TQh_d)$. Here, the matrix $Q$ is used to handle the irreflexive property of the directed interactions and can be fixed as an identity matrix for undirected interactions. To keep the embedding space almost isotropic, we expect $Q$ to be far away from singularity and thus sample the components of $Q$ iid from $\mathcal{N}(0,1)$ (line 1 in \proj)~\cite{tao2012topics}. The above parameterization is crucial because it guarantees that the embedding space indeed reflects structures of real networks where nodes from the same group may share similar patterns. We illustrate this point via Prop.~\ref{prop}. See its proof in Supplement~\ref{app:prop1}~\cite{CodeAndSupplement}. \vspace{-0.5mm}

\begin{proposition} \label{prop}
Suppose a node $v$'s embedding $h_v$ lies in the convex hull of the embeddings of a group of nodes $C= \{v_1, v_2,...,v_k\}$, i.e. $h_v = \sum_{i=1}^ka_ih_{v_i}$ for some non-negative $\{a_i\}_{i=1}^k$ such that $\sum_{i=1}^k a_i =1$. If for all $(s,d)$, $f_{sd}$ follows a Gamma distribution $\mathbb{P}(f;\lambda_{sd})$ and $\lambda_{sd} = \exp(h_s^TQh_d)$, then for any node $u\in V$, both $\mathbb{E}[f_{uv}]$ and $\mathbb{E}[f_{vu}]$ are controlled via $\min_{1\leq i\leq k} \mathbb{E}[f_{uv_{i}}] \leq \mathbb{E}[f_{uv}] \leq \sum_{i=1}^k a_i \mathbb{E}[f_{uv_{i}}]$ and $ \min_{1\leq i\leq k} \mathbb{E}[f_{v_{i}u}] \leq \mathbb{E}[f_{vu}] \leq \sum_{i=1}^k a_i \mathbb{E}[f_{v_{i}u}]$ respectively.
\vspace{-1mm}
\end{proposition}


Our factorization approach utilizes the maximum-likelihood rule to calculate node embeddings based on $F$. Recall that node embeddings are collected in $H = \{h_v|v\in V(F)\}$ and $\lambda_{vu} = \exp(h_v^TQh_u)$. The node embeddings $H$ can be estimated via:
\begin{align}
    \max_{H}\sum_{\langle(s,d),(t,f)\rangle\in F} \log \mathbb{P}(f;\lambda_{sd}) +  \sum_{(s',d')\in F^c}\log \mathbb{P}(f_{\text{th}};\lambda_{s'd'}). \label{eq:ffobj}
   \vspace{-0.5mm}
\end{align}
The first term of Eq.~\eqref{eq:ffobj} consists of high frequency interaction types ($>f_{\text{th}}$), while the second term consists of other interaction types ($F^c\triangleq[V(F)]^2\backslash \text{ITS}(F)$) that are either infrequent ($\leq f_{\text{th}}$) or have never even appeared. The second term is necessary because the ITFM $F$ only records the approximation of the sparse network structure to satisfy the memory constraint. Even if $F$ would record all the interactions that have appeared, we find in our experiments that a small positive $f_\text{th}$ improves the robustness of the model. Moreover, in practice, as $V(F)$ could be large, one may use mini-batch stochastic gradient ascent to optimize Eq.~\eqref{eq:ffobj}, where the first term may be sampled from $F$ and the second term is sampled from interaction types not included in $F$.

\begin{figure}[t]
\includegraphics[trim=0.9cm 0.5cm 0.5cm 0.1cm, clip, width=0.15\textwidth]{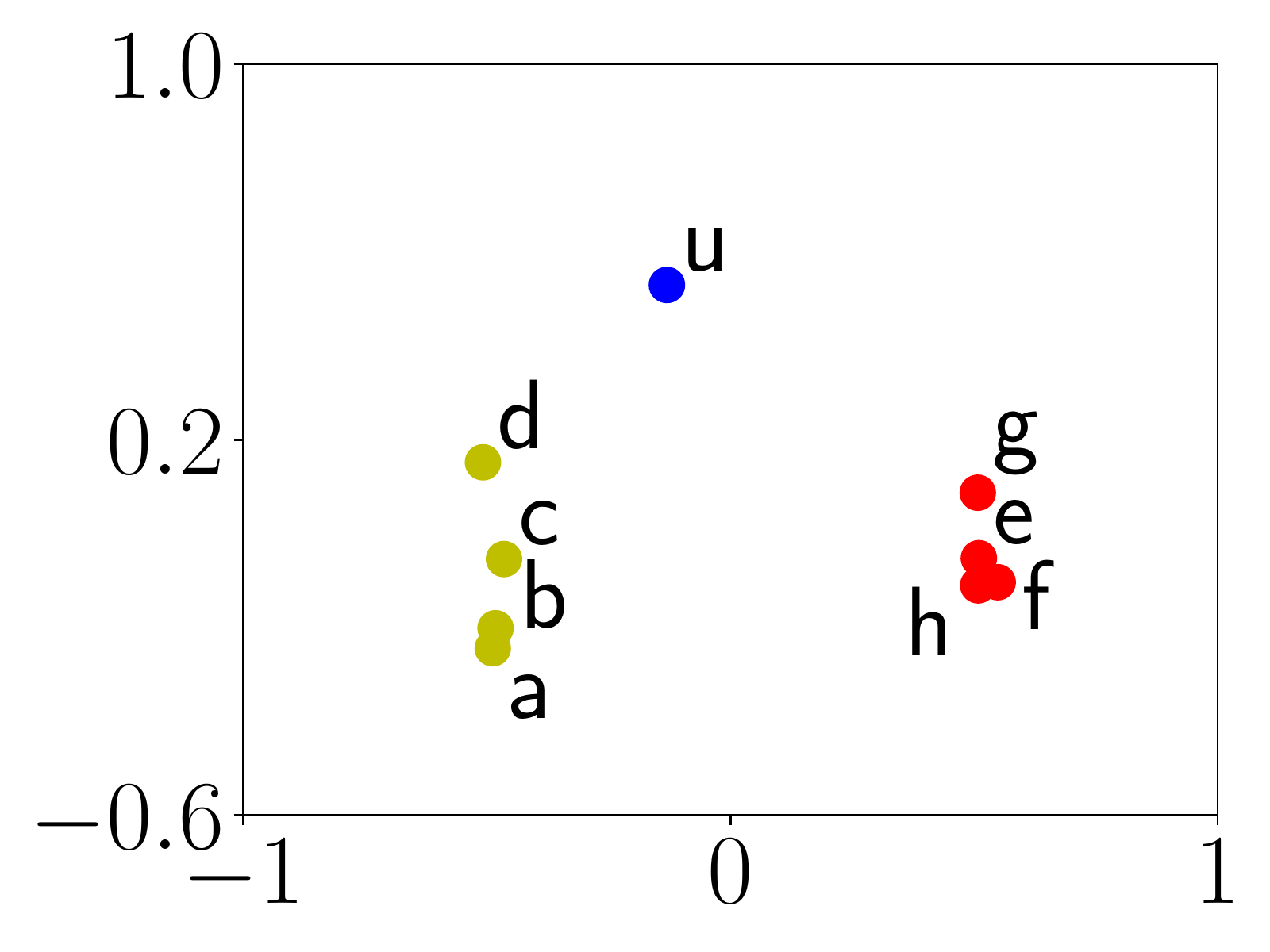}
\includegraphics[trim=0.9cm 0.5cm 0.5cm 0.1cm, clip, width=0.15\textwidth]{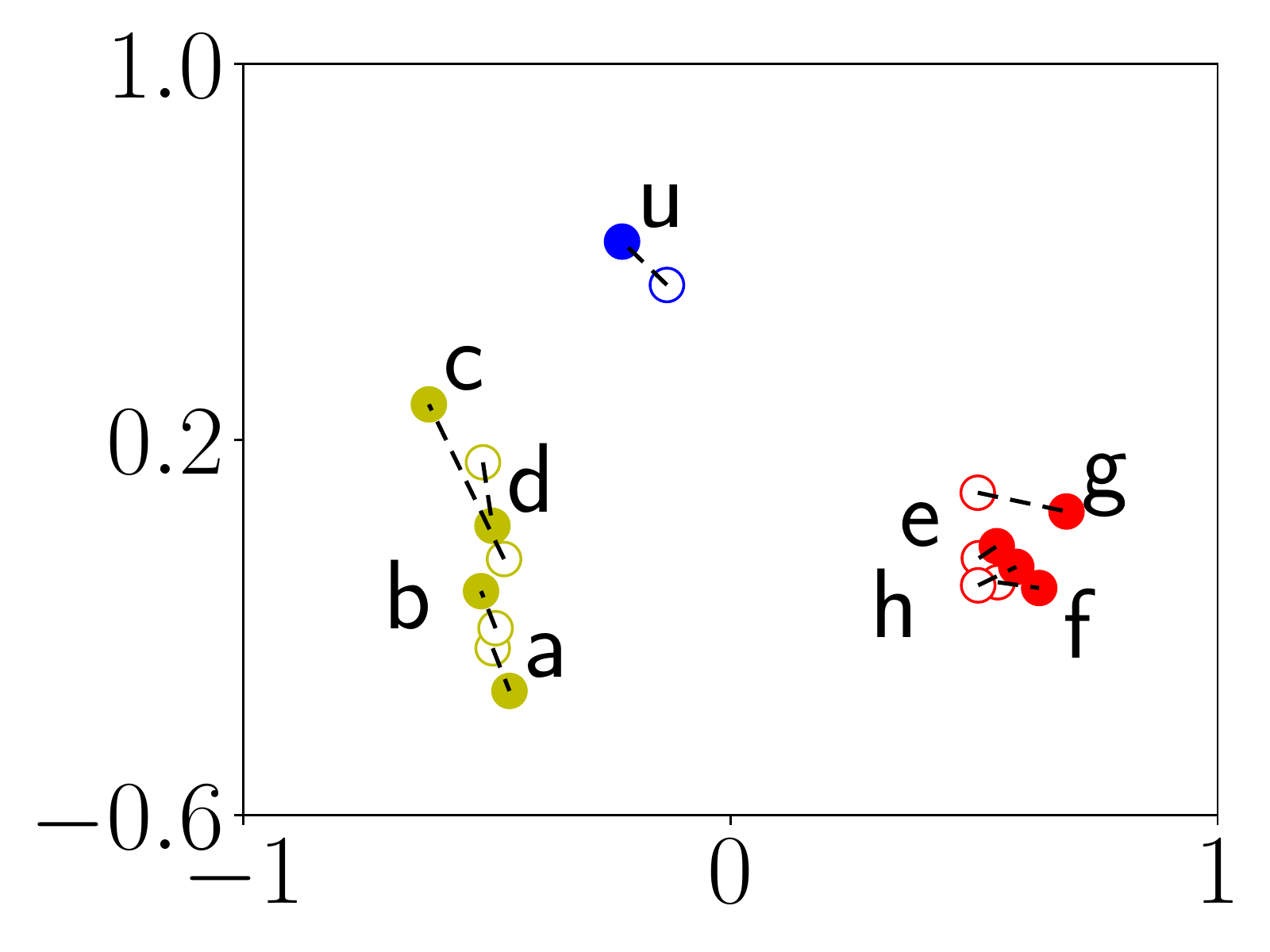}
\includegraphics[trim=0.9cm 0.5cm 0.5cm 0.1cm, clip, width=0.15\textwidth]{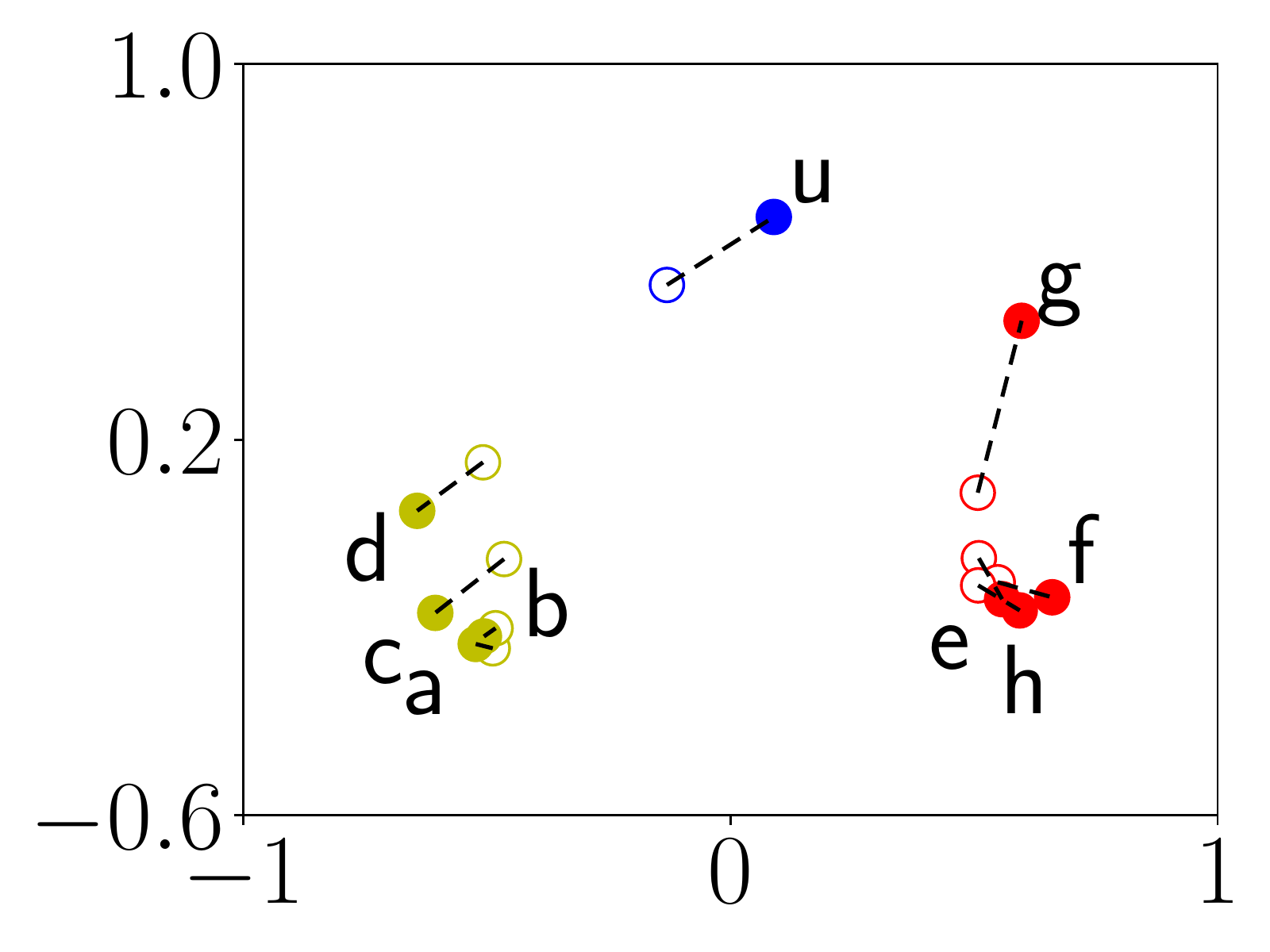}
\vspace{-2mm}
\caption{Visualizations of the movement of node embeddings learnt via \proj when interaction pattern changes from Fig.\ref{fig:pattern} (i) (Left) to Fig.\ref{fig:pattern} (ii) (Mid.) or (iii) (Right) respectively. Assign the regular freq. as 5 and $f_{\text th}$ as 0.005.} 
\label{fig:node-emb}
\vspace{-3.5mm}
\end{figure}

\vspace{-0.5mm}
\xhdr{Online Model Update} At $t=t_{\text{setup}}$, the network skeleton is available in $F$ and we can optimize the embeddings of all the nodes recorded in $V(F)$, which is the same as $V(\text{Act-S})$. For $t=t_{\text{setup}} + kW_{\text{upd}}$, $k\geq 1$, we update the model in an online fashion by decreasing the computation complexity. Note that $\text{Act-S}$ records the types of interactions that appear in the most recent update window. As the time-evolving frequencies recorded in $F$ may significantly change only for the types of interactions in $\text{Act-S}$, F-FAC focuses on optimizing the embeddings of nodes in $V(\text{Act-S})$ that is a subset of $V(F)$. We summarize the whole procedure into F-FAC (Alg.~\ref{alg:ffac}). Visualizations of the learnt node embeddings of patterns in Fig.\ref{fig:pattern} are shown in Fig.\ref{fig:node-emb}. We may see that the group change can be reflected via the movement of node embeddings. 

\vspace{-0.8mm}
\subsection{Online Detection}\label{sec:detect}
\vspace{-0.2mm}
Next, we consider how to assign the anomaly score for each interaction, i.e., the DETECT subroutine (line 4 of \proj), summarized in Alg.\ref{alg:detect}. In the previous subsection, we introduced \proj to learn the parameters of distributions of frequencies. The anomaly score of each interaction depends on the likelihood of the observed frequency of this interaction with respect to the learned distribution. We define \emph{the observed frequency} of one interaction as follows.
\begin{definition}
\emph{The observed frequency} of an interaction is defined as the inverse of the time difference between the time of this interaction and the previous appearance of the same-type interaction. 
\end{definition}

Let $f_{sd}^{(o)}$ denote the observed frequency of one $(s,d)$-type interaction. Its likelihood, based on our model for the distribution of the regular frequency, is computed as $\text{lh}(f_{sd}^{(o)}) = \mathbb{P}(f_{sd}^{(o)};\lambda_{sd})$ where $\mathbb{P}(\cdot)$ follows Eq.~\eqref{eq:f-model}. In general, by following the Neyman-Pearson lemma~\cite{neyman1933ix}, we can identify interactions with low likelihood values as anomalous. However, the values of $\text{lh}(f_{sd}^{(o)})$ may not be directly comparable because their underlying distributions hold different parameters. A further calibration is needed which can be accomplished as follows. We set the anomaly score of an interaction as the negative log probability to observe a frequency that follows the same distribution and has a lower likelihood value, i.e.,
\begin{align}\label{eq:anomaly-scores}
    \text{Sc}(f_{sd}^{(o)})  \triangleq - \log \mathbb{P}[\text{lh}(f)\leq \text{lh}(f_{sd}^{(o)})] = f_{sd}^{(o)}/\lambda_{sd},
\end{align}
where $f$ denotes a random variable that follows exactly $\mathbb{P}(f;\lambda_{sd})$ (Eq.~\eqref{eq:f-model}). As $H$ only tracks active nodes in $V(F)$, we may not have embeddings for nodes $s,\,d$ (including new arriving nodes and nodes with less frequent interactions) and thus set $\lambda_{sd}$ in Eq.~\eqref{eq:anomaly-scores} as: \vspace{-1mm}
\begin{align}
    \lambda_{sd} = \left\{\begin{array}{lc} \exp(h_s^TQh_d) & \text{if $h_s,h_d\in H$} \\ f_{\text{th}} & \text{otherwise} \end{array}\right.
\end{align}
This setting of anomaly scores promises control on the false positive rate if the model fits the distributions of the regular frequencies, as proved in Prop.~\ref{prop:false-pos}, and thus makes anomaly scores calibrated and comparable.  See the proof of Prop.~\ref{prop:false-pos} in Supplement~\ref{app:prop2}~\cite{CodeAndSupplement}.\vspace{-1.5mm}
\begin{proposition}\label{prop:false-pos}
If the model $\mathbb{P}(f;\lambda)$ (Eq.~\eqref{eq:f-model}) matches the distribution of the regular frequency and an interaction is determined as anomaly, because its anomalous score (set as Eq.~\eqref{eq:anomaly-scores}) is greater than a threshold $\tau$, then the obtained false positive rate is $\exp(-\tau)$. 
\end{proposition}

 \vspace{-2mm}
\xhdr{Computation of Observed Frequencies} 
Suppose we observe interactions $e=(s,d,t,w)$. The ITFM $F$ may contain the time $t'$  when the $(s,d)$-type interaction occurred previously.
Conversely, $F$ may not contain the time $t'$, if the $(s,d)$-type interaction has never occurred before or occurred a long time ago.

\begin{figure}
\includegraphics[trim=2.1cm 10.7cm 5.4cm 4cm, clip, width=\columnwidth]{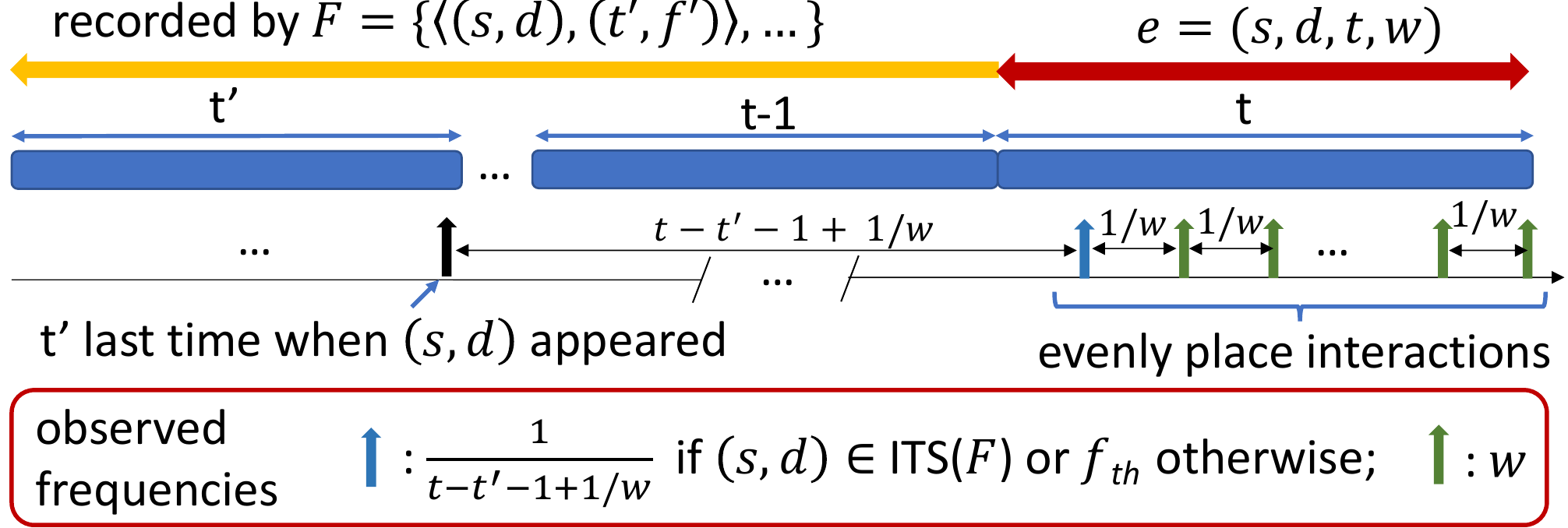}
\vspace{-6.5mm}
\caption{Computation of observed frequencies for $(s,d)$-type interactions at time $t$. }
\label{fig:ob-freq}
\vspace{-3mm}
\end{figure}

If $w=1$, the observed frequency of this single interaction is simply $1/(t-t')$ if ($(s,d)$ is in ITS(F) or $f_{\text{th}}$ otherwise. If several same-type interactions occur simultaneously, i.e. $w>1$, we place these $w$ interactions evenly within the time slot $t$, i.e., at $\{t-1+1/w, t-1+2/w,...,t\}$.
Based on this assumption, the observed frequency of the last $w-1$ $(s,d)$-type interactions is exactly $w$ while that of the first $(s,d)$-type interactions, according to the definition, is $1/(t-t'-1+1/w)$ if $(s,d)$ is in ITS(F) or $f_{\text{th}}$ otherwise. Fig.~\ref{fig:ob-freq} illustrates the above computation of observed frequencies.

\vspace{-1mm}
\xhdr{Group-level Detection} So far, the discussion has been concerned with anomalies of single-type interactions. However, as shown in pattern Fig.~\ref{fig:pattern} (v), some anomalies may only be detected when we simultaneously consider a group of interactions with different types. Our approach can be easily generalized to such group-based patterns.
We can combine interaction types of interest in a group of interaction types $\Xi =\{(s_1,d_1),...,(s_k,d_k)\}$, view this group as one type, and then
compute the observed frequencies with the same method that we use for single-type interactions.
Suppose the observed frequency for one interaction with type in $\Xi$ is $f_{\Xi}$, then the anomaly score is computed as: \vspace{-0.6mm}
\begin{align}\label{eq:anomaly-scores-group}
    \text{Sc}(f_{\Xi}) \triangleq f_{\Xi}/\sum_{(s,d)\in\Xi}\lambda_{sd} = f_{\Xi}/\sum_{(s,d)\in\Xi}\exp(h_s^TQh_d) \vspace{-0.6mm}
\end{align}
where $\sum_{(s,d)\in\Xi}\lambda_{sd}$ denotes the intensity of this group of interactions, which replaces the single $\lambda_{sd}$ in Eq.~\eqref{eq:anomaly-scores}. The intuition behind the use of the sum operation in Eq.~\eqref{eq:anomaly-scores-group} comes from the fact that single-type interactions whose arriving times satisfy an exponential distribution essentially correspond to a Poisson process and that merging multiple independent Poisson processes yields another Poisson process with the intensity that equals the sum of intensities of individual processes~\cite{poissonlecture}. Although any group $\Xi$ can be investigated, in this work, we focus on the group of interactions that share common source nodes or common destination nodes.


\begin{algorithm}[t]
\SetKwInOut{Input}{Input}\SetKwInOut{Output}{Output}
\Input{An ITFM $F$, $e$; Param. $H$, $Q$, $f_{\text{th}}$}
\Output{The anomaly score for each interaction in $e$} 
$\Xi_{\text{out},s} \leftarrow \{(s,d')|(s,d') \text{ occurs simultaneously at $t(e)$}\}$\;
$\Xi_{\text{in},d} \leftarrow \{(s',d)|(s',d) \text{ occurs simultaneously at $t(e)$}\}$\;
Compute the observed frequency of each of these $w(e)$ interactions according to their own type $(s,d)$, the group types $\Xi_{\text{out},s}$ and $\Xi_{\text{in},d}$, denoted by $f_{sd}$, $f_{\Xi_{\text{out},s}}$, $f_{\Xi_{\text{in},d}}$ respectively  \; 
Compute $\text{Sc}(f_{sd})$, $\text{Sc}(f_{\Xi_{\text{out},s}})$, $\text{Sc}(f_{\Xi_{\text{in},d}})$ based on Eqs.~\eqref{eq:anomaly-scores},~\eqref{eq:anomaly-scores-group} and~\eqref{eq:anomaly-scores-group} respectively, and output the anomaly score as $\max\{\text{Sc}(f_{sd}), \text{Sc}(f_{\Xi_{\text{out},s}}), \text{Sc}(f_{\Xi_{\text{in},d}})\}$  
\caption{DETECT ($F$, $e$, $H$, $Q$, $f_{\text{th}}$)}\label{alg:detect}
\end{algorithm}
\begin{table}[t]
\vspace{-10mm}
\end{table}


 \vspace{-0.8mm}
\subsection{Complexity Analysis and Discussion}\label{sec:comp}
 \vspace{-0.2mm}
The \emph{memory cost} of \proj is determined by parameter $M$ which controls the size of ITFM $F$. The sizes of $\text{Act-S}$ and $H$, according to their definitions, depend on the size of $F$, and are no greater than one time and two times of this size, respectively. 

The analysis of the \emph{online time complexity} of \proj is more complicated. Two operations are computationally demanding, the maintenance of the ITFM $F$ and an online update of $H$. $F$ requires efficient operations including search via keys, insert, and delete operations (the subroutine UNION) and thus should be implemented via a hash map. 
The most complex operation over $F$ is finding the minimum frequency (line 4 of UNION). A min heap which tracks the frequencies recorded in $F$ can perform this operation within $\log(M)$. Overall, the time complexity for $F$ is O($\log(M)$) per interaction.

Considering online updates of $H$, although each step could be complex, a constant number of epochs of gradient ascent
typically yields accurate enough solutions (10 epochs in our experiments in Sec.~\ref{sec:exp}). 
Our model further benefits from the fact that the product of matrices $H^TQH$ may be computed in parallel by parallel computing units such as GPUs, which further significantly improves the efficiency. The mini-batch training of F-FAC also allows to accommodate potential memory limits on a GPU. In our implementation, we also find it empirically unnecessary to traverse all $u\in V(F)$. Traversing all $u\in \mathbf{N}(v, F)$ with a few nodes sampled from $V(F)\backslash \mathbf{N}(v, F)$ (negative examples) has achieved good performance. 



\vspace{-1mm}
\section{Experiments} \label{sec:exp}
\vspace{-0.2mm}

\begin{table}[t]
\centering
\begin{tabular}{c|cccc}
\hline
Dataset & Nodes & Edges & Year & \# of Anomalies\\
\hline
\hline
BARRA1 &  38,408 & 1.64M & 12/2013 - 03/2020 & 5,856\\
BARRA2 &  49,189 & 2.22M & 06/2011 - 03/2020 & 1,255\\
BARRA3 &  63,209 & 2.35M & 09/2012 - 03/2020 & 33\\
BARRA4 &  138,940 & 5.60M & 09/2009 - 03/2020 & 8\\
\hline
\end{tabular}
\caption{Statistics of BARRA1-4}
\label{tab:BARRACUDA statistics}
\vspace{-9mm}
\end{table}

\begin{table*}[t]
    \centering
    \begin{tabular}{c|c|c|c|c|c|c|c}
    \hline
    Methods & RTM-InjectionS & RTM-InjectionW & DARPA & BARRA1 & BARR2 & BARRA3 & BARRA4\\
    \hline
    \hline
    SedanSpot & 0.521 $\pm$ 0.012 & 0.472 $\pm$ 0.059 & 0.657 $\pm$ 0.004 & 0.427 $\pm$ 0.059 & 0.414 $\pm$ 0.172 & 0.524 $\pm$ 0.099 & 0.679 $\pm$ 0.024\\
    AnomRank & 0.553 & 0.549 & 0.764 & $0.837^*$ & $0.731^*$ & N/A & N/A\\
    NetWalk & 0.516 $\pm$ 0.022 & 0.599 $\pm$ 0.013 & 0.732 $\pm$ 0.033 & N/A & N/A & N/A & N/A\\
    Midas & $\textbf{0.958}^*$ & $0.993^*$ & $\textbf{0.947}^*$ & 0.559 & 0.563 & 0.733 & 0.446\\
    F-FADE & 0.719 $\pm$ 0.012 & \textbf{1.000 $\pm$ 0.000} & 0.920 $\pm$ 0.004 & \textbf{0.875 $\pm$ 0.001} & \textbf{0.822 $\pm$ 0.007} & \textbf{0.781 $\pm$ 0.018} & \textbf{0.941 $\pm$ 0.012}\\
    \hline
    \end{tabular}
    \caption{Anomaly detection performance comparison in AUC (mean $\pm$ $95\%$ confidence level for randomized algorithms). * highlights the best baselines. Bold fonts highlight the optimal performance among all methods. N/A indicates that the methods cannot make one pass of those BARRA1-4 datasets within 10 hours with 10 minutes as the systemic time granularity.}
    \label{tab:performance}
    \vspace{-6mm}
\end{table*}

In this section, we evaluate the performance of \proj compared to state-of-the-art anomaly detection methods on dynamic graphs. We focus on answering three questions. \textbf{Q1. Accuracy:} How accurately does \proj detect synthetic and real-world anomalies with labels compared to the baselines? \textbf{Q2. Effectiveness:} Is \proj able to detect meaningful real-world events? \textbf{Q3. Parameter Sensitivity:} As discussed in Sec.\ref{sec:comp}, the entire memory cost and time complexity essentially depend on the size of $F$, i.e., $M$. Therefore, how does $M$ affect the detection performance?

\hide{
\todo{Yen-Yu ToDo}
\pan{Thanks! This is great. Let me know if you encounter any questions.}
\begin{itemize}
    \item Rebuild our pipeline
    \item Test $\alpha$ decay function
    \item Fix $Q$
    \item DBLP dataset
    \item Enron email dataset special events figure
    \item Embedding visualization (v.s. MIDAS, matrix factorization)
\end{itemize}
}
\vspace{-1mm}
\subsection{Experimental Setup}
\vspace{-1mm}

\xhdr{Datasets} We use one synthetic and seven real-world networks, where the obtained anomalies can be verified by comparing them to manual annotations or by correlating them with real-world events.

\emph{RTM}~\cite{4781165} refers to a model to generate synthetic weighted time-evolving graphs based on Kronecker products~\cite{leskovec2010kronecker}, which successfully matches several of the properties of real graphs. We follow the same setting in~\cite{yoon2019fast} to generate a directed graph with 1K nodes and 8.1K directed edges over 2.7K timestamps. As our input data is edge streams, we randomly permute all edges with the same timestamp and merge them into the stream. We further inject two types of anomalies to evaluate different models. \emph{InjectionS:} 
At each of 50 randomly selected timestamp, we randomly choose 8 nodes, inject all edges between them in both directions. 
\emph{InjectionW:} We uniformly at random select 50 timestamps. At each each of 50 randomly selected timestamp, inject 70 simultaneous edges between two randomly selected nodes. 
InjectionS and InjectionW mimic patterns Fig.\ref{fig:pattern}~(v) and (iv) respectively.

\emph{DARPA}~\cite{lippmann1999results} is a network traffic dataset simulating various intrusion behaviors. It contains 4.5 M IP-IP communications (directed edges) taking place between 9,484 source IPs and 23,398 destination IPs (nodes) over 87.7K minutes. Anomalous communications are associated with labels that can be used for evaluation. 


\emph{ENRON}~\cite{shetty2004enron} has 50K emails (directed) exchanged among 151 employees (nodes) over 3 years (from 01/1999 to 06/2002) in ENRON Corporation. Since there are no labels to represent whether an email is anomalous or not, we apply \proj to detect the event of a sudden increase in email communication among the employees. 

\emph{DBLP}\footnote{https://dblp.uni-trier.de/xml/} is the collaboration graph of authors from the DBLP computer science bibliography. The nodes in this graph represent the authors, and edges between two authors represent joint publications. For simplicity, we focus on the papers published in 1960-2010. Overall, we obtain a graph with around 653K nodes and 2.9M edges.  Note that this dataset is undirected so we choose $Q = I$ in our model. There are no available labels for anomalies and instead we expect \proj to detect unlikely collaborations, e.g., authors suddenly changed in their coauthorship activities. 

\emph{BARRA datasets}\footnote{https://www.barracuda.com/. Ethic claim: 
Authorized employees at Barracuda were allowed to access the data (under standard, strict access control policies). No personally identifying information or sensitive data was shared with any non-employee of \\Barracuda. 
Once Barracuda deployed a set of ATO detectors to production, any detected attacks were reported to customers in real time.
} include the email networks sampled over about the past decade used in multiple organizations who are customers of Barracuda Networks. Barracuda Networks is a large security company that focuses on providing developed anomaly detection solutions over commercial email systems for multiple organizations. Data is provided in the form of sender, recipient, timestamp, and label. Sender and recipient are the hashed email addresses of emails' senders and recipients. 
The label field indicates whether an email is an Account Takeover (ATO) attack or not, which is in-prior obtained via Barracuda internal evaluation. We choose four email networks related to four organizations denoted as \emph{BARRA1,2,3,4} respectively with their overview in Table \ref{tab:BARRACUDA statistics}. Relevant research related to phishing detection over this data has been published~\cite{ho2019detecting}, where a supervised learning method based on email content features was proposed. However \proj is purely unsupervised, does not require access to the email content and thus offers better privacy protections.

\xhdr{Baseline} We choose four most recently proposed baselines for comparison including \emph{SedanSpot}~\cite{eswaran2018sedanspot}, \emph{Midas}~\cite{bhatia2020midas}, \emph{AnomRank}~\cite{yoon2019fast} and \emph{NetWalk}~\cite{yu2018netwalk} and  use the implementations provided by their authors\footnote{SedanSpot: https://github.com/dhivyaeswaran/sedanspot;
Midas: https://github.com /Stream-AD/MIDAS;
AnomRank:https://github.com/minjiyoon/KDD19-AnomRank; NetWalk: https://github.com/chengw07/NetWalk.}. Note that SedanSpot~\cite{eswaran2018sedanspot} and Midas~\cite{bhatia2020midas} are the SOTA distance-based and probabilistic methods, respectively, to detect anomalies in edge streams. We are not aware of any matrix factorization based methods for detecting anomalies in edge streams. Therefore, we further consider AnomRank~\cite{yoon2019fast} and NetWalk~\cite{yu2018netwalk} which were proposed for graph streams. As they are not directly applied to edge streams, we properly revise them to make a fair comparison. They are both fed with the graph streams aggregated from edges in each time window under the finest time granularity. AnomRank~\cite{yoon2019fast} computes PageRank scores of different nodes and NetWalk~\cite{yu2018netwalk} tracks node embeddings. Both methods provide anomaly scores for each node that are related to at least one edge in the current time window. We associate an edge with the anomaly score equal to the greater one of its two corresponding end-nodes. 





\begin{figure*}[t]
\begin{minipage}[b]{0.23\linewidth}
\centering
\small 
\begin{tabular}{c|cccc}
\hline
Dataset & $\alpha$ & $M$ & $m$ & initial $f_{\text{th}}$\\
\hline
\hline
DARPA & 0.999 & 200 & 100 & 16.7\\
ENRON & 0.999 & $\infty$ &100 & 0\\
DBLP & 0.999 & $\infty$ & 100 & 0\\
BARRA1 & 0.999 & 100 & 200 & 2.6\\
BARRA2 & 0.999 & 400 & 200 & 0.93\\
BARRA3 & 0.999 & 400 & 200 & 1.2\\
BARRA4 & 0.999 & 400 & 200 & 1.1\\
\hline
\end{tabular}
\vspace{-3mm}
\caption{Hyper-parameters of F-FADE, the unit of $f_{\text{th}}$ is $10^{-3} \text{minute}^{-1}$.}
\label{tab:hyper-parameters}
\end{minipage}\hspace{0.005\linewidth}
\begin{minipage}[b]{0.76\linewidth}
\vspace{-3mm}
\flushright

\subfloat{
\raisebox{0.3in}{\rotatebox[origin=t]{90}{\textbf{ENRON}}}
\includegraphics[trim=0cm 0.1cm 0.9cm 0.1cm, clip,width=0.85\textwidth]{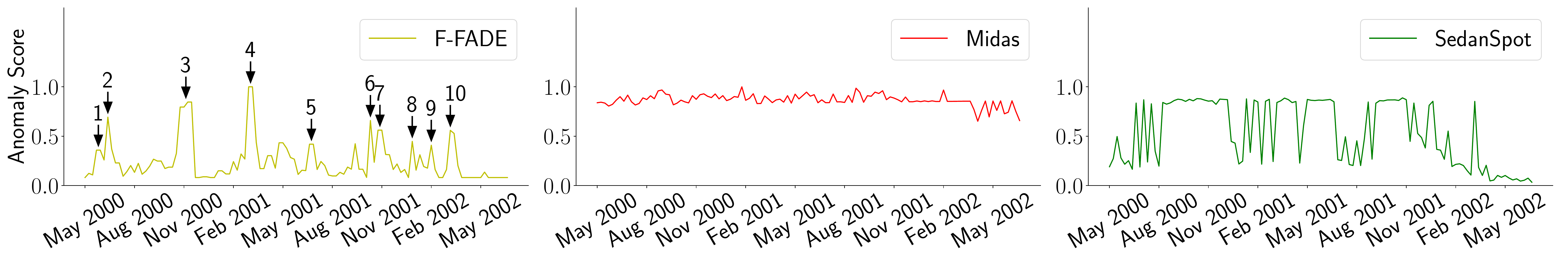}
}\\
\subfloat{
\raisebox{0.3in}{\rotatebox[origin=t]{90}{\textbf{DBLP}}}
\includegraphics[trim=0cm 0.1cm 0.9cm 0.1cm, clip,width=0.85\textwidth]{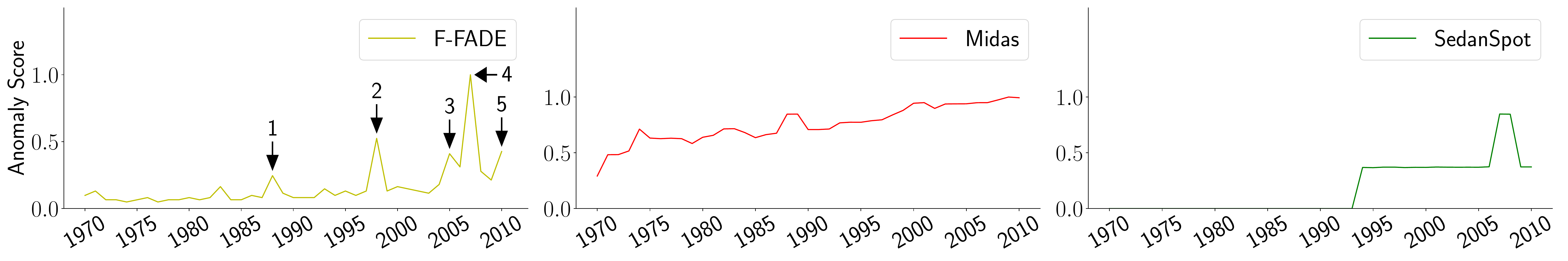}
}
\vspace{-3mm}
\caption{Real-world events detection over ENRON and DBLP networks.}
\label{anomaly_detection}
\end{minipage}
\end{figure*}

\xhdr{Evaluation}
For the RTM graph, DARPA, and BARRA1-4, which have labeled anomalies, we use the area under curve (AUC) score to measure the anomaly detection performance of all methods. We set up all the models based on the data in the first $10\%$ of total time, and evaluate them over the rest of data in the remaining $90\%$ of time. For randomized algorithms including SedanSpot~\cite{eswaran2018sedanspot}, NetWalk~\cite{yu2018netwalk} and \proj, their corresponding AUC scores are summarized based on 10 randomly independent tests. For ENRON and DBLP, as they do not have labels, we evaluate the methods by correlating the predicted anomalies with real-world events. 
For ENRON, we set up the models on data from 01/1999 - 04/2000 and use 05/2000 - 06/2002 for evaluation. For DBLP, we set up the models on data from the years 1960-1969, and use the years 1970-2010 for evaluation. We performed hyper-parameter tuning for all baselines and report the best performance. For \proj, we show the hyper-parameters in Table~\ref{tab:hyper-parameters} and properly tune $W_{upd}$ to report the best performance. Recall $M$ is the size of the ITMF $F$, which further relates to both the space and time complexity of \proj~ (Sec. \ref{sec:comp}). We will further investigate its effect in Sec.~\ref{sec:parameter sensitivity}. More details of experimental settings are described in Supplement~\ref{app:exp}~\cite{CodeAndSupplement}. \vspace{-1mm}

\vspace{-1mm}
\subsection{Accuracy of Anomaly Detection}
\vspace{-0.2mm}

We used AUC scores to evaluate the performance of all methods over the RTM, DARPA and BARRA datasets (see Table~\ref{tab:performance}). 
SedanSpot performs poorly on the RTM-InjectionS and RTM-InjectionW tasks, because patterns (iv) and (v) can hardly be detected based on the changes in the personalized PageRank scores that SedanSpot essentially tries to approximate.
AnomRank also performs poorly on those tasks, with a lower performance than shown in the original paper~\cite{yoon2019fast}, because it is applied to detecting edge-level anomalies which requires a more timely response.
AnomRank exhibits sensitivity to the choice of the time granularity. This issue can be observed for AnomRank also on the DARPA dataset. However, AnomRank performs well on the BARRA1 ad BARRA2 networks because the variation of interaction frequencies in the email networks seems to be smaller than that in computer network traffic (DARPA) and thus the choice of the time-window length for AnomRank is not that critical. However, AnomRank is not able to process two largest networks BARRA3 and BARRA4 on time. NetWalk performs relatively well on the DARPA network but is not able to process the four large BARRA networks. Midas performs very well over the RTM and DARPA datasets, which demonstrates the benefit of probabilistic models to control false positive rates. Midas is the best performing method for the DARPA dataset as the anomalies mostly consist of the patterns (iv) and (v) in Fig.~\ref{fig:pattern}. However, this dataset does not contain communities or other low-rank structures, which are present in real social networks, including the four BARRA networks. Midas cannot leverage these network structures and 
performs worse than \proj over the BARRA datasets. In contrast, \proj can control false positive rates and can handle all of these patterns. Therefore, \proj performs uniformly well over all datasets.

\vspace{-0.5mm}
\subsection{Effectiveness in Detecting Events}\label{Effectiveness}
\vspace{-0.2mm}
Here, we focus on the three methods that are proposed to process edge streams, \proj, Midas and SedanSpot, and evaluate their capabilities to detect social events over ENRON and DBLP. In order to visualize the results, we aggregate edges occurring in each week on ENRON by taking their max anomaly scores per week, and in each year on DBLP by taking their max anomaly score per year. Note that these two datasets were previously used to evaluate SedanSpot~\cite{eswaran2018sedanspot}, where the events are also aggregated weekly for ENRON and yearly for DBLP. However, an additional threshold was defined to determine whether an event is anomalous and the number of anomalous events per week or per year are used to make the event detection. We do not use the evaluation in~\cite{eswaran2018sedanspot} because it may introduce three extra hyper-parameters that are challenging to tune as the anomaly scores provided by different methods are not on the same scale. Our evaluation does not depend on extra hyper-parameters and is thus more equitable. 


Over ENRON, F-FADE and SedanSpot show some similar trends, but SedanSpot outputs many high scores unrelated to any true events. For Midas, most of the output scores are high and without much correlation with interesting events. The anomalies detected by F-FADE coincide with major events annotated as (1)-(10) in the ENRON timeline. We explain these events in Supplement~\ref{app:enron}~\cite{CodeAndSupplement}.

Over DBLP, Midas does not perform well. 
SedanSpot works better than Midas and still outputs many similar max anomaly scores. In contrast,  F-FADE effectively detects many temporal anomalous coauthorships that are annotated as (1)-(5). Among these events, SedanSpot only detects (4). We verify the anomalies (1)-(5) using the public profiles of the authors and list them in Supplement~\ref{app:dblp}~\cite{CodeAndSupplement}.






\hide{
\begin{textblock*}{5cm}(6.8cm,2.9cm) 
   \textbf{ENRON}
\end{textblock*}

\begin{textblock*}{5cm}(7cm,4.8cm) 
  \textbf{DBLP}
\end{textblock*}
}

\vspace{-0.5mm}
\subsection{Parameter Sensitivity}\label{sec:parameter sensitivity}
\begin{figure}[t]
\includegraphics[width=0.49\textwidth, clip]{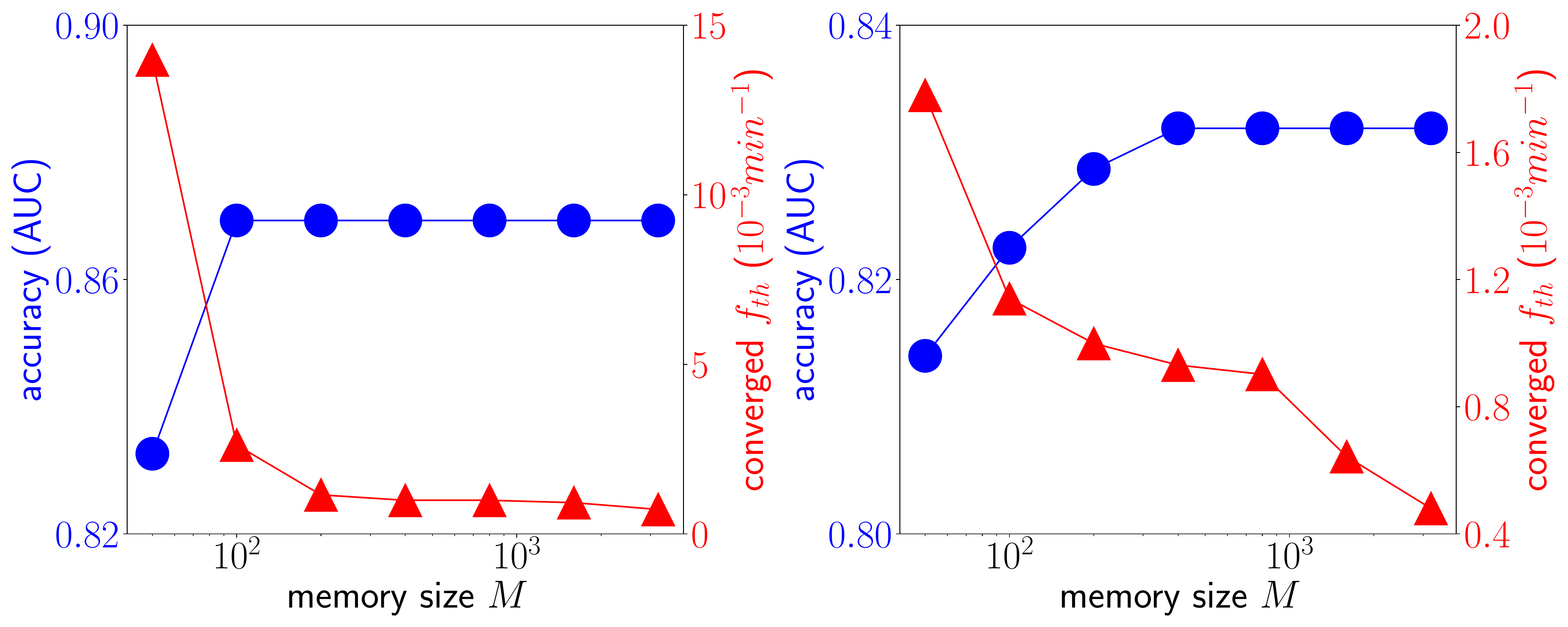}
\vspace{-8mm}
\caption{The AUC scores of anomaly detection and the convergent $f_{\text{th}}$'s of \proj OVER BARRA1 (left) and BARRA2 (right) with respect to different memory size $M$'s.}
\label{fig:parameter sensitivity}
\vspace{-3mm}
\end{figure}

\vspace{-0.2mm}
As we analyzed in Sec.~\ref{sec:comp}, the memory size $M$ of $F$ is the key parameter that determines both the space and time complexity of F-FADE. $M$ may also affect the anomaly detection performance of \proj via $f_{\text{th}}$: $f_{\text{th}}$, as the cut-off frequency, determines the aggregated frequency of any node-pairs of which the network skeleton $F$ may loss track and the new arriving node-pairs. Given an initial cut-off frequency $f_{\text{th}}$, $M$ directly impacts how $f_{\text{th}}$ varies when \proj runs through the edge streams. Therefore, we would like to understand how $M$ affects $f_{\text{th}}$ and further affects the performance of \proj. We evaluate \proj over BARRA1 and BARRA2 with different values of $M$ changing from 50 to 3,200 with power of 2. The results are summarized in Fig.~\ref{fig:parameter sensitivity}. As expected, a larger $M$ may lead to a smaller convergent $f_{\text{th}}$, which implies that $F$ may track a boarder range of interaction frequencies and thus the performance of \proj improves. Simultaneously, a greater $M$ introduces higher memory and time costs. Interestingly, rather small $M$'s (100 for BARRA1 or 400 for BARRA2) have already achieved almost the optimal performance, although both datasets contain a large number of edges, 1M+ and 2M+ respectively, which means that \proj works well at relatively small memory and time costs. We suspect that the regularization based on $f_\text{th}$ in Eq.~\eqref{eq:ffobj} and the network structures help greatly.





\section{Conclusion and Future Work}
In this work, we propose \proj that is a purely unsupervised, online approach for detecting anomalies in edge streams. \proj takes advantage of both probabilistic models and matrix factorization by factorizing time-evolving intensities of interaction frequencies. \proj provides false-positive-rate guarantees in the detection of a single or a group of anomalous edges. 
\proj incurs only a constant memory cost by recording a network skeleton that consists of the most frequent interactions. Extensive experiments demonstrate the power of \proj to effectively capture temporal and structural changes. Finally, the success of \proj raises many interesting directions for future studies including investigating the node embedding space obtained from frequency factorization, developing new approaches to combine the frequency-factorization techniques with network attributes to detect anomalies, and designing an automatic mechanism of maintaining the memory cost to enlarge the parameter regime that is optimal for \proj.

\begin{acks}
We thank Phil Porras for discussions and suggestions.
We also gratefully acknowledge the support of
DARPA under No. FA865018C7880 (ASED),
ARO under No. W911NF-16-1-0342 (MURI),
NSF under No. OAC-1835598 (CINES), CCF-1918940 (Expeditions),
and Stanford Data Science Initiative.
\end{acks}

\bibliographystyle{ACM-Reference-Format}
\balance
\bibliography{acmart.bib}
\newpage
\appendix
\section{Proof of Proposition~\ref{prop}} \label{app:prop1}

Since $f_{sd} \sim \mathbb{P}(f;\lambda_{sd})$ where $ \mathbb{P}(f;\lambda_{sd})$ is a Gamma distribution that follows $\mathbb{P}(f;\lambda) = \frac{1}{\lambda^\theta \Gamma(\theta)}f^{\theta-1}\exp(-\frac{f}{\lambda})$, $\mathbb{E}[f_{uv}]$ can be represented as $\mathbb{E}[f_{uv}] = \theta\lambda_{uv}$. Recall that we parameterize $\lambda_{sd} = \exp(h_s^TQh_d)$ for all $(s.d)$ pairs. We suppose that $h_v$ lies in the convex hull of the embeddings of a group of nodes $C= \{v_1, v_2,...,v_k\}$, i.e., $h_v \sum_{i=1}^k a_i h_{v_i}$ for non-negative and $\ell$-1 normalized $\{a_i\}_{i=1}^{k}$. Then, the upper bound of $\mathbb{E}[f_{uv}]$ can be derived as follows:

\begin{align*}
    &\frac{\mathbb{E}[f_{uv}]}{\theta} =\lambda_{uv}= \exp(h_u^TQh_v) = \exp\left(h_u^TQ(\sum_{i = 1}^k a_i h_{v_i})\right) \\
    &\leq \sum_{i = 1}^k a_i \exp(h_u^TQh_{v_i})= \sum_{i = 1}^k a_i \lambda_{uv_i}= \sum_{i = 1}^k a_i \frac{\mathbb{E}[f_{uv_i}]}{\theta},
\end{align*}
where the inequality is due to the convexity of the exponential function.
Regarding the lower bound of $\mathbb{E}[f_{uv}]$, we can derive it as follows:

\begin{align*}
    &\frac{\mathbb{E}[f_{uv}]}{\theta} = \lambda_{uv} = \exp(h_s^TQh_d) = \exp\left(h_u^TQ(\sum_{i = 1}^k a_i h_{v_i})\right)\\
    &= \exp\left(\sum_{i = 1}^k a_i (h_u^TQh_{v_i})\right) \geq \exp\left(\sum_{i = 1}^k a_i \min_{1\leq i\leq k} h_u^TQh_{v_i}\right)\\
    &= \exp\left(\min_{1\leq i\leq k} h_u^TQh_{v_i}\right) = \min_{1\leq i\leq k} \exp(h_u^TQh_{v_i}) = \min_{1\leq i\leq k} \lambda_{uv_i}\\
    &= \min_{1\leq i\leq k} \frac{\mathbb{E}[f_{uv_i}]}{\theta},
\end{align*}
where the inequality is due to non-negativity of $\{a_i\}_{i=1}^{k}$ and we also used $\sum_{i=1}^k a_i =1$.




By combining the upper bound and the lower bound of $\mathbb{E}[f_{uv}]$, we prove that $\mathbb{E}[f_{uv}]$ is controlled by $\min_{1\leq i\leq k} \mathbb{E}[f_{uv_{i}}] \leq \mathbb{E}[f_{uv}] \leq \sum_{i=1}^k a_i \mathbb{E}[f_{uv_{i}}]$. The inequality of $\mathbb{E}[f_{vu}]$ can be derived similarly in the same way.

\section{Proof of Proposition~\ref{prop:false-pos}}\label{app:prop2}
As for the assumption, $\mathbb{P}(f;\lambda)$ (Eq.~\eqref{eq:f-model}) matches the distribution of the regular frequency. Then, for an interaction that is not anomaly, its observed frequency should follow $f^{(o)}\sim \mathbb{P}(f;\lambda)$. Our method will detect it as anomaly if, according to Eq.~\eqref{eq:anomaly-scores},
\begin{align*}
    \frac{f^{(o)}}{\lambda} \geq \tau.
\end{align*}
Then, the false positive rate is nothing but the probability such that the above inequality is satisfied. That is, when $f^{(o)}\sim \mathbb{P}(f;\lambda)$,
\begin{align*}
    \mathbb{P}\left(\frac{f^{(o)}}{\lambda} \geq \tau\right) = \mathbb{P}\left( f^{(o)} \geq \lambda\tau\right) = \exp\left( -\frac{\lambda\tau}{\lambda}\right) = \exp\left(-\tau\right),
\end{align*}
which concludes the proof.

\section{Training Configuration} \label{app:exp} 
We performed hyper parameter search for best performance for our method and all the baselines and used the following hyper-parameters to obtain the reported results:

\begin{table*}[ht]
\centering
\begin{tabular}{c||c|c|c|c|c}
\hline
Dataset & $W_{\text{upd}}$ & $\alpha$ & $M$ & $m$ & initial $f_{\text{th}}$\\
\hline
\hline
RTM-InjectionS & 5, 10, 20, 40, 80, 160 & 0.999 & 100, 300, 500, 700, 1000 & 25, 50, 100 & 0.77\\
RTM-InjectionW & 5, 10, 20, 40, 80, 160 & 0.999 & 100, 300, 500, 700, 1000 & 25, 50, 100 & 3.13\\
DARPA & 60, 120, 360, 720, 1440 & 0.999 & 100, 200, 500, 1000, 2000 & 100, 150, 200 & 16.7\\
ENRON & 10080 & 0.999 & $\infty$ & 100 & 0\\
DBLP & 1 & 0.999 & $\infty$ & $100$ & 0\\
BARRA1 & 10080, 21600, 43200 & 0.999 & 100, 200, 400, 800, 1600, 3200 & 100, 150, 200 & 2.6\\
BARRA2 & 10080, 21600, 43200 & 0.999 & 100, 200, 400, 800, 1600, 3200 & 100, 150, 200 & 0.93\\
BARRA3 & 10080, 21600, 43200 & 0.999 & 100, 200, 400, 800, 1600, 3200 & 100, 150, 200 & 1.2\\
BARRA4 & 10080, 21600, 43200 & 0.999 & 100, 200, 400, 800, 1600, 3200 & 100, 150, 200 & 1.1\\
\hline
\end{tabular}
\caption{Hyperparameters and their value for F-FADE on different dataset}
\label{tab:F-FADE hyperparameter}
\vspace{-9mm}
\end{table*}

For RTM graph, DARPA, and BARRA1-4, we setup all the models based on the data in the first $10\%$ total time. Table~\ref{tab:hyper-parameters} lists the hyperparamters and their values. The unit of $W_{\text{upd}}$ is year for DBLP and minute for others, and the unit of initial $f_{\text{th}}$ is $10^{-3} \text{minute}^{-1}$.

\hide{
\textbf{RTM:} I need to check the training records on workstation. I will put the statistics later

\textbf{DARPA:} $W_{\text{upd}} = \{60, 120, 360, 720, 1440\}$ minutes, $\alpha = 0.999$, $M = \{100, 500, 1000, 2000\}$, $m = \{100, 150, 200\}$ initialize $f_{\text{th}}$ as $0.0167$ $\text{minute}^{-1}$.

\textbf{BARRACUDA:} $W_{\text{upd}} = \{10080, 21600, 43200\}$ minutes, $\alpha = 0.999$, $M = \{100, 200, 400, 800, 1600, 3200\}$,  and initialize $f_{\text{th}}$ as (2.6, 0.93, 1.2, 1.1) $\times 10^{-3} \text{minute}^{-1}$ for (BARRA1, BARRA2, BARRA3, BARRA4).

\textbf{ENRON:} $t_{\text{setup}} = 692K$ minutes, $W_{\text{upd}} = 10080$ minutes, $\alpha = 0.999$, $M = \infty$, $m = 100$, and initialize $f_{\text{th}}$ as 0.

\textbf{DBLP:} $t_{\text{setup}} = 10$ years, $W_{\text{upd}} = 1$ year, $\alpha = 0.999$, $M = \infty$, $m = 100$, and initialize $f_{\text{th}}$ as 0.
\pan{Please reorganize: what unit is $f_th$. All methods require four parameters $\alpha, M, m, $initial $f_{th}$. Moreover, a table for this may be more clear. Where is the hyperparameter for the synthetic data? Also, for randomized algorithms (Netwalk and our method), you should do multiple-time experiments and show the statistics of these methods. }
}

For baselines, we used the implementations provided by their authors and we report the range of configurations used for baselines here:

\textbf{SedanSpot:} $num walks = \{5, 10, 20, 50, 100\}$, $restart\_prob = \{0.05, 0.1, 0.15, 0.2, 0.5\}$, $sample\_size = \{50, 100, 200, 500, 1000\}$ on synthetic graphs, and $sample\_size = \{10K, 20K, 50K\}$ on DARPA and BARRACUDA, and following hyper-parameter settings as suggested in the original paper on ENRON and DBLP.

\textbf{AnomRank:} $aggregation\_timesteps = \{10, 30, 60, 180, 360, 720,\\ 1440\}$ minutes on synthetic graphs, DARPA, and BARRACUDA.

\textbf{NetWalk:} $representation\_size = \{20, 50, 100\}$, $num\_walks = \{2, 3, 5\}$, $walk\_length = \{3, 5, 10\}$, $\rho = \{0.1, 0.2, 0.3\}$, $k = \{5, 10, 20\}$, $\lambda = \{0.0005, 0.001, 0.005\}$, $\beta = \{0.1, 0.2, 0.5\}$, $\gamma = \{1, 5, 10\}$, $\alpha = \{0.3, 0.5, 0.7\}$, $snap\_size = \{500, 1000, 2000\}$ for synthetic graphs. $embedding\_size = \{20, 50, 100\}$, $\alpha = \{0.3, 0.5, 0.7\}$, $k = \{5, 10, 20\}$, $snap\_size = \{250K, 500K\}$, and following the other hyper parameter settings as suggested in the original paper for real-word graphs on DARPA. $learning\_rate = 0.01$ for adam optimizer as suggested in the public source code.

\textbf{Midas:} $decay\_factor = \{0.3, 0.5, 0.7\}$, $num\_hash = \{2, 5, 10\}$, $num\_buckets = \{500, 1000, 2000, 5000\}$ on RTM graph, DARPA, and BARRACUDA. $decay\_factor = 0.5$, $num\_hash = 10$, and $num\_buckets = 5000$ for ENRON and DBLP.

\section{Events detection in ENRON}\label{app:enron}
In this section we demonstrate the effectiveness of F-FADE on ENRON dataset in the main paper \ref{Effectiveness}. The anomalies detected by F-FADE coincide with major events in the ENRON timeline \footnote{https://www.agsm.edu.au/bobm/teaching/BE/Enron/timeline.html} as follows:
\begin{enumerate}
    \item 05/22/2000: The California ISO (Independent System Operator), the organization in charge of California's electricity supply and demand, declares a Stage One Emergency, warning of low power reserves.
    \item 06/12/2000: Skilling makes joke at Las Vegas conference, comparing California to the \textit{Titanic}.
    \item 11/01/2000: FERC investigation exonerates Enron for any wrongdoing in California.
    \item 03/2001: Enron transfers large portions of EES business into wholesale to hide EES losses.
    \item 07/13/2001: Skilling announces desire to resign to Lay. Lay asks Skilling to take the weekend and think it over.
    \item 10/17/2001: Wall Street Journal article reveals the details of Fastow's partnerships and shows the precarious nature of Enron's business.
    \item 11/08/2001: Enron files documents with SEC revising its financial statements for past five years to account for $586$ million in losses. The company starts negotiations to sell itself to Dynegy, a smaller rival, to head off bankruptcy,
    \item 01/25/2002: Cliff Baxter, former Enron vice chairman, commits suicide.
    \item 02/02/2002: The Powers Report, a 218-page summary of an internal investigation into Enron's collapse led by University of Texas School of Law Dean William Powers, spreads blame among self-dealing executives and negligent directors.
    \item 03/14/2002: Former Enron auditor Arthur Andersen LLP indicted for obstruction of justice for destroying tons of Enron-related documents as the SEC began investigating the energy company's finances in October 2001.
\end{enumerate}

\section{Events detection in DBLP}\label{app:dblp} 
In this section we demonstrate the effectiveness of F-FADE on DBLP dataset in the main paper \ref{Effectiveness}. We expect anomalous edges to represent unlikely collaborations. We verify anomalies using the public profiles of the authors as follows:
\begin{enumerate}
    \item 1988: G. M. Lathrop and J. M. Lalouel have 15 coauthor papers, but they don’t have any coauthor paper before.
    \item 1998: Raj Jain has 63 papers this year, and he has 32 coauthor papers with Rohit Goyal and Sonia Fahmy. But Raj Jain has only 5 papers in 1997, and 4 of them are coauthor papers with Rohit Goyal and Sonia Fahmy.
    \item 2005: Elizabeth Dykstra-Erickson and Jonothan Arnowitz have 25 coauthor papers in this year. But before 2005, they have only 1 coauthor paper in 2003.
    \item 2007: Damien Chablat and Philippe Wenger have 61 coauthor papers, but they only have 1 coauthor paper in 2006.
    \item 2010: Alan Dearle and Graham N. C. Kirby have 27 coauthor papers. But before 2010, they have most 3 coauthor papers in 2003.
\end{enumerate}

\end{document}